# Metallization of hydrogen

M. I. Eremets *, P. P. Kong, A. P. Drozdov

*Max-Planck Institute for Chemistry; Hahn-Meitner Weg 1, Mainz 55128, Germany*

**Abstract**
In previous work,[1] we showed that hydrogen metallizes in phase III at temperatures below ~200 K and at pressures near ~350 GPa. Here, we perform a detailed study of electrical conductivity R(T) in phase III over a pressure range of 200-400 GPa and a temperature range of 80-300 K, and we show that hydrogen transforms from a semiconducting to a metallic state already at ~315 GPa. This transformation is also supported by Raman spectroscopy: the Raman signal intensity decreases with pressure in accordance with the appearance and increase of electrical conductivity. Moreover, the Raman and electrical measurements yield the same boundary between the hydrogen phases III and V.

**Introduction**.

Metallic hydrogen is a subject of very long intensive studies since the prediction of Wigner and Huntington[2] (1935) that insulating molecular hydrogen will dissociate at high densities and the atomic hydrogen will be an alkali-like metallic substance. Metallic hydrogen should be high temperature, likely room temperature superconductor[3]. Moreover, metallic hydrogen is predicted to be liquid at low temperatures and this liquid hydrogen, a double-degenerate Fermi system formed by electrons and protons is predicted to be superconducting superfluid[4-6]. Recent works predict the Wigner and Huntington[2] metallization in the range of 500 GPa in the most stable atomic hydrogen candidate structure has space group *I*41/*amd* [7-9]. In particular, McMinis et al [9] using quantum Monte Carlo calculations found that molecular phase *C*2/*c* is stable almost up to the molecular to atomic transition at pressure of 447(3) GPa.

Wigner and Huntington[2] considered a dramatic transformation in hydrogen – dissociation from molecule to atomic state which should be metallic. However, hydrogen can turn to metal very differently, in a molecular state, before the eventual dissociation[1,10,11]. With pressure, the band gap of a molecular dielectric phase can gradually decrease and finally collapse. As the result of the bands overlap free carriers appear: electrons in the conduction band and holes in the valence band - hydrogen turns to metal. In the initial state of the overlap, the concentration of the carriers is small and hydrogen is a poor metal with low conductivity similar to semimetals such as bismuth. With pressure, the band overlap increases, and conductivity increases so that this semimetal gradually transforms to good metal. This way of metallization follows from the recent calculations[10,11].

Experimentally, metallization of hydrogen was claimed several times (see review[12]). The experiments on metallization are very difficult, performed at ultimate pressures, and typically this is not a systematic study. Recent statements[13,14] are based mostly on optical measurements. Dias and Silvera[13] concluded that hydrogen metallized at 495 GPa based on the visual observation (reflecting sample) and measurements of reflection at two wavelengths. This work was severely criticized[15-19], and it was not reproduced. In the recent experiment on infrared absorption,[14] transition to metallic state was evidenced from a sudden loss of the transmission of light at a pressure above 425 GPa, which was assigned to a structural transition from insulating *C*2/*c*-24 structure to a metallic *Cmca*-12 structure and abrupt closure of the bandgap. However the absorption measurements cannot directly prove the metallic zero-gap state as they were done down to 800 cm$^{-1}$, and therefore a possibility of a small gap $E_g < 0.1$ eV remains. This experiment was not repeated so far or confirmed by other methods.

We consistently work on the metallization of hydrogen with electrical measurements[1,20,21]. This method is indispensable for direct proving a metallic state, in which electrons are free, and metal conducts to the lowest temperatures, in contrast to a semiconductor, which also conducts but turns to an insulator at sufficiently low temperature as it has an energy band gap.

We observed first a metallic state in Ref.[20]. At cooling at ~360 GPa from a semiconducting phase IV, the resistance strongly dropped, and at lower temperatures, it was nearly temperature independent. We interpreted this temperature dependence as an evidence of a metal, however, the absolute value of resistance was high in comparison with typical metals.

In the next work[1] we focused on the low temperature part of the phase diagram (phase III) where the metallic state was discovered and performed a more systematic study[20]. We found that hydrogen was insulating in phase III at low pressures, but at pressures above ~350-360 GPa conductivity emerged, and we measured the temperature dependence R(T) at several pressures up to 440 GPa. In accordance with our previous measurements[20], we observed weak temperature dependence of resistance, which absolute value was higher than that for typical metals. The perfect Raman spectra, which we measured up to 480 GPa gave a key for understanding the metallic state. It turned out that the Raman signal persisted in the metallic state while its intensity dropped at pressures above ~330 GPa. The Raman spectra eventually vanished at 440 GPa. The energy of phonon and vibron bands in the Raman spectra smoothly changed with pressure, which showed that there was no major structural transition between 200-440 GPa; this means that the transition of the metallic state is not structural but an electronic transformation. The persistence of the Raman spectra in the metallic state shows that hydrogen is a poor metal at least up to 440 GPa, and likely it transforms to a good molecular or atomic metal at higher pressures.

We concluded that the combination of the metallic temperature behaviour of the high resistance hydrogen together with the persistence of the Raman signal could be explained by semimetallic, poor metal state of hydrogen. This conclusion was supported by similar examples of the pressure-induced metallization in xenon, oxygen, and bromine[1].

Subsequent state-of-art theoretical calculations[10,11] showed that molecular hydrogen in phase III with *C*2/*c* structure should turn to metal through the closure of the band gap. The calculations naturally explained the low optical absorption and persistence of the Raman signal at the metallization by the low concentration of free carriers (electrons and holes) at the beginning of the overlapping of the conductive and valence bands. A significant absorption was predicted only at pressures ~460 GPa (optical gap closure)[11] in agreement with the vanishing of the Raman signal at pressure ~440 GPa[1].

Despite the conclusive results and the agreement with the theory, further study of metallic hydrogen is obviously desirable. A drawback of the previous studies[1,20] is that the electrical measurements were done only in few pressure points, and the pressure of transition from the dielectric to metallic states in phase III was not accurately determined. In addition, the temperature range of the metallic state was not well defined.

In the present work, we further confirmed the metallization of hydrogen[1,20] and extended the previous electrical conductivity study to the domain of 200-404 GPa and 80-290 K. We established metallic hydrogen in phase III at a pressure above ~315-320 GPa where hydrogen transforms from semiconducting to semimetallic state. This transition coincides with the onset of the decrease of the Raman intensity with pressure.

**Experimental**

For the electrical measurements, we sputtered at an anvil four Ta electrodes of ~70 nm thick and covered them by ~70 nm thick layer of Au. The electrodes were sputtered on the ~50 nm thick layer of $Al_2O_3$ deposited on the diamond surface. The electrodes were insulated from the metallic gasket (T301 stainless steel) with cBN/epoxy, MgO/epoxy, CaO/epoxy layers. We clamped hydrogen in DAC with the aid of a gas loader at a pressure of ~0.1 GPa, and then increased pressure to ~100 GPa, cooled the cell down to 180 K, and then increased pressure at low temperature. We increased load at low temperature for a precaution, to prevent failure of diamond, which frequently happens in phase IV at room temperature in the 200-300 GPa range. During the gradual increase of pressure and the Raman measurements, the temperature raised, and at pressures about 325-330 GPa hydrogen transformed from phase III to phase IV as clearly seen from the drastic change of the Raman spectra (SM Fig.1).

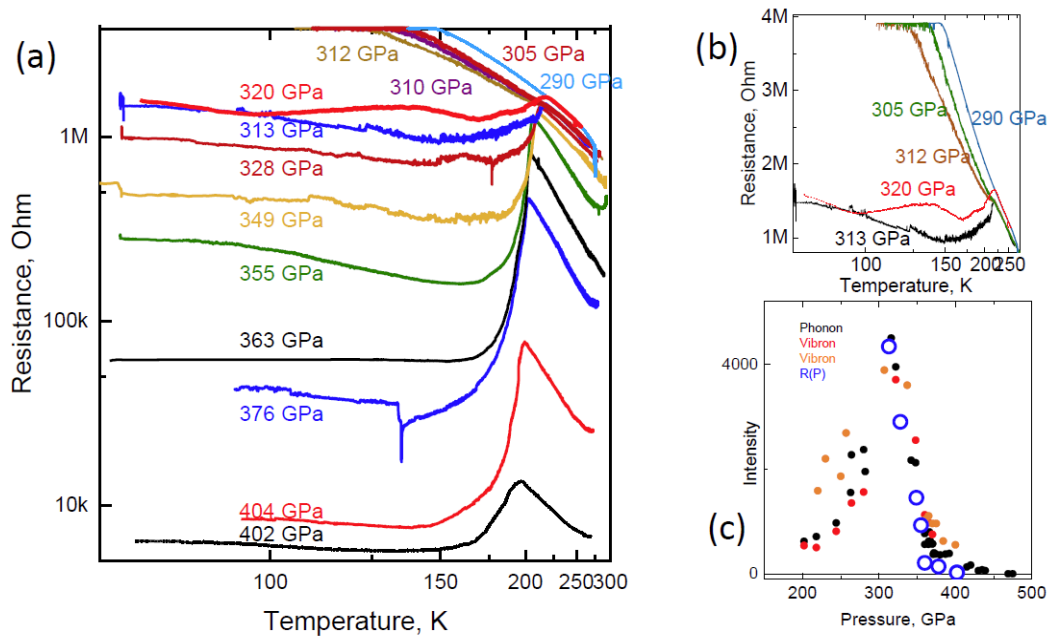

Fig. 1. The temperature dependence of resistance R(T) of hydrogen at different pressures. (a) The R(T) measurements were done at a decrease of pressure down to 290 GPa. The plots at the lowest pressures are shown in detail in (b). Note semimetallic behavior at 313 GPa, but at lowering pressure to 312 GPa and 305 GPa, it changes to semiconducting one. At 290 GPa the resistance of the sample becomes too high and the measured resistance comes likely from a spurious residual resistance of the insulating layer of $Al_2O_3$ or the gasket CaO/epoxy. The saturation of the resistance at the upper parts of the curves is determined by the limit of the measurements of ~3.9 Mohm. The subsequent increase of pressure from 290 to 320 GPa restores the semimetallic state (red thick curve). (c) Comparison between the pressure dependence of resistance with the Raman intensity. The resistance was taken at 100 K at different pressures (blue circles). The rest points were intensities of the Raman lines in the spectra of hydrogen at 100 K taken from Ref. 1, Fig. 2f.

We started measurements of the temperature dependence of resistance at the highest pressure of 402 GPa and then gradually, with small steps decreased pressure and measured R(T) at each pressure point (Fig. 1). At 402 GPa, the resistance was measured with four probes first, but soon one electrode became unstable and therefore we continued the measurements with two electrodes (SM Fig. 2). In this case, the resistance of the electrodes ~10 kOhm is added to the resistance of the sample but likely it is significantly smaller than the resistance of the sample at least at low pressure range where it is in the MOhm range (Fig. 1). At cooling, the resistance R(T) strongly dropped at 197 K and then rose very slowly with a small activation energy of ~0-1 meV (Fig. 1). This small increase of resistance with cooling is characteristic for semimetals[22]. The point of inflection at 197 K (the maximum at the R(T)) coincides with the boundary of phases V and III at the phase diagram (Fig. 2). This is not a random coincidence at the particular pressure of 402 GPa but the maximum at the R(T) follows the boundary between phases III and V at other pressures too (Fig. 2).

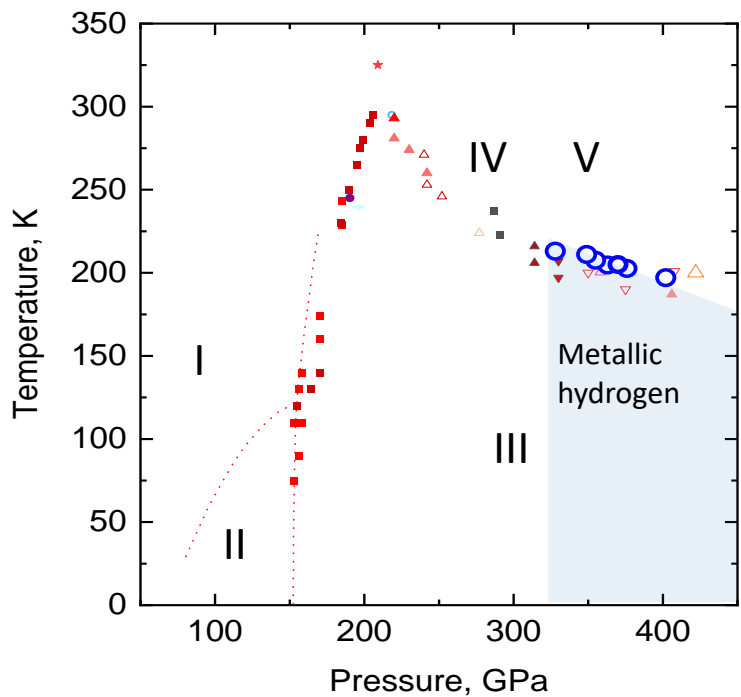

Fig. 2. The phase diagram of hydrogen. The blue open circles are the points of transition between phases III and IV or V derived from the electrical measurements. The rest experimental points are obtained in Ref [21] from the changes of the Raman spectra at crossing lines between different phases (I-V). Dot lines are phase boundaries from Ref. [24].

With the decrease of pressure (Fig. 1), the value of resistance of hydrogen increases but the temperature dependence R(T) remains to be semimetallic. Finally, at decreasing of pressure below ~315 GPa it changes to semiconducting (Fig 1b, curves at 312 GPa and 305 GPa), and then to insulating behaviour. This change is reversible: after increasing pressure from 290 GPa to 320 GPa the insulating hydrogen restores to semimetal. Thus, we specified the pressure of transformation of hydrogen to the metallic state in phase III at ~315-320 GPa (Fig. 2).

### Discussion

We confirmed metallic hydrogen discovered in the previous works[1,20]. A new finding is that metallic hydrogen occupies phase III at pressures above ~315 GPa, it persists at least to 440 GPa as shown in Ref. [1] and to 402 GPa in the present work. It lies at temperatures below phases IV and V down at least to 80 K and likely at lower temperatures as the metallic behavior was determined down to 25 K in Ref. [20] at 360 GPa.

We also established the boundary between phases V and III with the aid of electrical measurements. This boundary coincides with the boundary, which is well established in the Raman studies (Fig. 2). The Raman spectra sharply change[20] (SM Fig. 1) at crossing the phase boundary between two very different structures: $C2/c$ in phase III and $Pc$ in phase IV[23] or similar structure $Pca2_1$ in phase V at pressure above ~ 275 GPa[24]. The structures were proposed from the theoretical calculations and they are likely correct as agree with the Raman spectroscopy: the boundary between phases IV (V) and III was established with Raman spectroscopy up to 370 GPa[20] and we extended it up to 420 GPa (Fig. 2).

The exact determination of the boundary of the metallic state at ~315 GPa allowed us to give a new interpretation of the Raman data. The dramatic drop of the Raman signal at pressures above ~330 GPa observed in Ref. [1] (Fig.1c) was difficult to connect with the closure of the indirect gap, which was defined at a higher pressure of ~350-360 GPa. Instead, it was hypothesized that the peak in Raman intensity is connected with a resonance of the Raman scattering with the HeNe laser excitation (1.92 eV). In the present study, we found that the Raman spectra could be naturally connected with the transformation to the metallic state as the pressure of the onset of the metallic state at ~315-320 GPa is close to the pressure of the peak of the Raman intensity ~330 GPa (Fig.1c). Moreover, the resistance of the metallic hydrogen decreases with pressure in accord with the decrease of the Raman intensity (Fig. 1c). The pressures were determined with the aid of the diamond edge scale both for the electrical and Raman measurements. One should keep in mind an uncertainty of determination of pressure with the aid of the diamond edge scale[25]. It is based not on the measurements of the pressure inside the medium which surrounds the sample, but the stresses in the diamond, which is adjacent to the sample[25]. The pressure depends on the particular arrangement of DAC and can differ at up to ~20 GPa in different runs. The pressures from different runs nevertheless can be compared with the aid of hydrogen vibrons, which serves as an internal pressure gauge[20,26]. In Fig. 1c, where the pressures was determined with the diamond scale, the present electrical and the Raman measurements from Ref[1] (Fig. 2a) can be correctly compared as vibron frequencies measured for both runs well coincide (SM Fig. 1). In contrast, the pressure of 360 GPa determined in Ref. [1] as the emergence of the metallic state was overestimated at ~16 GPa in comparison of the pressures used in the present work (see an explanation in SM Fig. 3).

At higher pressures, above the transformation to semimetal, the decrease of the Raman intensity is naturally explained by the screening of the light due to the carriers, which appeared at the transformation to the metallic state. With pressure, the concentration of carriers increases, the Raman signal disappears at P>440 GPa, and one could expect that hydrogen turns into a good metal and probably a superconductor. Pressures up to 480 GPa were already explored with Raman spectroscopy[1].

Acknowledgements. M.E. is thankful to the Max Planck community for valuable support and and Prof. Dr. U. Pöschl for the encouragement.

**Supporting materials**

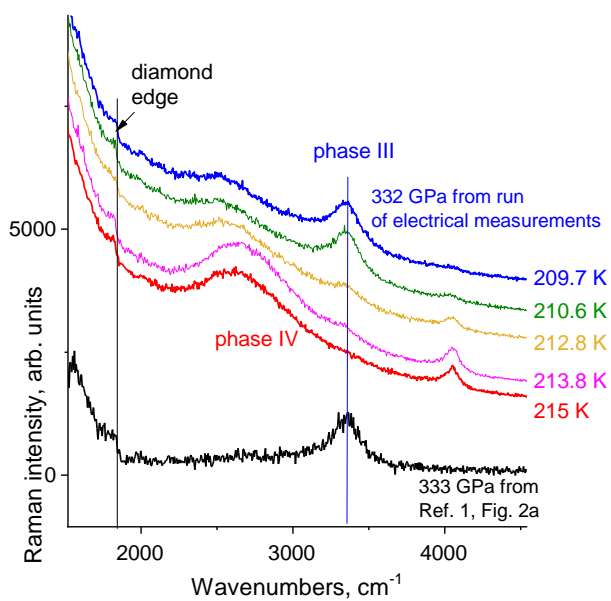

SM Fig. 1. Raman spectra of hydrogen at the transition across the boundary between phases III and IV at pressures ~330 GPa. At 209.7 K, the hydrogen vibron around 3350 cm$^{-1}$ corresponds to structure C2/c of the low temperature phase III. At 215 K, vibrons at ~2600 cm$^{-1}$ and ~4060 cm$^{-1}$ corresponds to the structure of the high temperature phase IV. In the intermediate range of temperatures, there is a mixture of phases III and IV. The black spectrum is shown for comparison. It is taken from Ref. [1] at the same pressure according to the diamond edge. The pressures are the same according to the same positions of the vibrons in phase III.

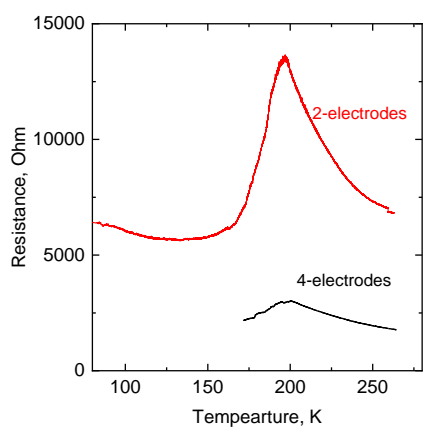

SM Fig. 2. The temperature measurements of resistance at 402 GPa with two and four probe measurements. At cooling from room temperature, the activation energy of $E_a$ ~20 meV deduced from R(T) indicates that hydrogen is a narrow band semiconductor.

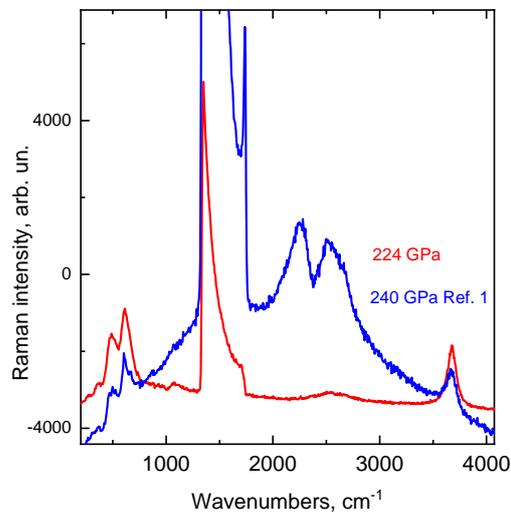

SM Fig. 3. The Raman spectrum from the present work and the spectrum from Ref. 1 (this particular spectrum is not plotted in Fig. 2a) are taken at the same pressure as follows from the same positions of the vibron, which serves as an internal pressure gauge. However, the pressures derived from the diamond edge differ, as shown in the plot. This means that the pressure of 360 GPa (emergence of the electrical conductivity) in Ref. 1, Fig. 3c is overestimated at ~16 GPa in comparison with the present work.